\documentclass[epj]{svjour}
\usepackage{setspace}
\usepackage[latin1]{inputenc}         
\usepackage[british]{babel}
\usepackage[T1]{fontenc}
\usepackage[normalem]{ulem}
\usepackage{color}
\usepackage{amssymb}
\usepackage{amsmath}
\usepackage{subfigure}

\usepackage{graphics}
\usepackage[all]{xy}  

\begin{document}
\title{Social dynamics with peer support on heterogeneous networks}
\subtitle{The "mafia model"}
\author{Marta Balb\'as Gambra \and Erwin Frey% etc
%\mail{
% \thanks is optional - remove next line if not needed
%\thanks{\emph{Present address:} Insert the address here if needed}%
}                     % Do not remove
\offprints{frey@lmu.de}          % Insert a name or remove this line
\institute{Arnold Sommerfeld Center for Theoretical Physics and CeNS,
  Department of Physics,\\ Ludwig-Maximilians-Universit\"at M\"{u}nchen,
  Theresienstrasse 37, D-80333 M\"{u}nchen, Germany}

\date{Received: date / Revised version: date}
% The correct dates will be entered by Springer
%
\abstract{Human behavior often exhibit a scheme in which individuals adopt indifferent, neutral, or radical positions on a given topic. The mechanisms leading to community formation are strongly related with social pressure and the topology of the contact network.  Here, we discuss an approach to model social behavior which accounts for the protection by alike peers proportional to their relative abundance in the closest neighborhood. We explore the ensuing non-linear dynamics emphasizing the role of the specific structure of the social network, modeled by scale-free graphs. We find that  both coexistence of opinions and consensus on the default position are possible stationary states of the model. In particular, we show how these states critically depend on the heterogeneity of the social  network and the specific distribution of external control elements.}
\PACS{
      {89.75Fb}{Structures and organization in complex systems} \and
      {64.60}{Phase transitions} \and
      {87.23}{Population dynamics} \and
      {87.10}{Stochastic modelling} 
     } % end of PACS codes
 %end of abstract
%
\maketitle
\section{Introduction}
Opinion and group formation in human societies have been investigated
in the last years as complex phenomena, well described with the
methods of non-linear dynamics and statistical physics (for a review
see \cite{castellano2009statistical}). It has become clear that the
evolutionary dynamics of 
societies, based on 
social pressure and imitation, must be understood together with the
intricate network of contacts established among single agents. Individuals
are represented by the nodes of a graph and interact with their
neighbors under a given set of rules to form their own
opinion. Interactions may take place in several ways, and many models
have dealt with various mechanisms in which individuals might meet as well as
with different rules to update their opinions. 

In the simplest \emph{voter model}~\cite{clifford1973model,holley1975ergodic,de1993non} individuals on the nodes of a regular lattice  may hold two opinions encoded by a binary spin variable  $\sigma=\pm 1$. Two randomly selected individuals interact with each other (\emph{one-to-one} interaction) in which the second simply adopts the opinion of the first. In one dimension, the voter model is identical to the kinetic Ising model with zero-temperature Glauber dynamics~\cite{liggett1985interacting,castellano2006zero,castellano2005comparison}. The latter  is
exactly solvable for regular lattices in any dimension~\cite{frachebourg1996exact,krapivsky1992kinetics}.
This simple dynamics leads to consensus in finite systems, which is
the only stable solution: all spins are aligned, while both states
are equally likely to be the consensus opinion---provided they are equally abundant in the initial state---i.e.\ ensemble magnetization
is conserved \cite{liggett1985interacting}.

On heterogeneous networks, special updating mechanisms like the \emph{link-update}~\cite{suchecki2005conservation,sood2008voter,sood2005voter} approach are needed to ensure conservation of the ensemble magnetization. For the \emph{reverse-voter}~\cite{castellano2005comparison,sood2008voter}, where a node is randomly selected and its opinion copied to a randomly selected neighbor, it was found that the time to consensus differs from the direct voter model. A common feature of the voter dynamics on various complex networks, such as in small-world~\cite{castellano2003incomplete}, scale-free~\cite{suchecki2005conservation,suchecki2005voter,sood2005voter}, and random~\cite{vazquez2008analytical} graphs, is the absence of consensus in the thermodynamic limit, which has also been supported by analytical investigations on uncorrelated
graphs~\cite{vazquez2008analytical}. However, consensus is systematically reached on finite graphs~\cite{sood2005voter,castellano2003incomplete}, even though metastable states of coexisting opinions are observed for long periods before the system evolves towards consensus for a time which scales with the system's size.

In the voter model, the inclusion of more than two opinions~\cite{sire1995coarsening,vazquez2003constrained}, non-confident vacillating voters~\cite{lambiotte2007dynamics}, zealots with a fanatic position~\cite{mobilia2007role,mobilia2003does}, or a threshold number of  successful encounters~\cite{dall2007effective} allows the
 coexistence of opinions. Encounters among many agents have been considered in the \emph{majority model}~\cite{krapivsky2001organization,chen2005consensus},
in which a group of individuals is randomly selected and all of  them adopt the opinion of the majority (\emph{all-to-all} interactions), or in the \emph{rumor spreading model}~\cite{galam2002minority,galam2007role}, in which more than one group-encounters take place simultaneously. In both of these models consensus on one or the other opinion depending on the initial conditions is achieved for various structures. Nevertheless, if noise is introduced in this group dynamics, as in the \emph{majority-minority}~\cite{mobilia2003majority} or the \emph{majority-vote}~\cite{liggett1985interacting} models, the system evolves  either towards a consensus or towards a state in which both opinions are equally represented (zero magnetization state in the Ising model analogy). The same behavior is observed in the \emph{Sznajd model}~\cite{sznajd2000opinion,slanina2003analytical} where two individuals sharing the same opinion impose it on their neighbors.

Another approach for describing social interactions that accounts for the influence of the whole neighborhood on a single individual (\emph{one-to-all} interactions) has been discussed in \emph{social impact theory}~\cite{latane1981psychology,latane1990private,lewenstein1992statistical}. An agent changes her opinion if the pressure in favour of the opinion change overcomes the support to keep the current position.  This model yields stable coexistence of opinions, unless the presence of external fields is taken into account, for which metastable states  are observed in which domains with the minority opinion successively shrink~\cite{lewenstein1992statistical}. 

The \emph{Abrams-Strogatz} (AS) model describes the competition between two languages in a population, with  non-linear transition rates proportional to $x^a$, where $x$ is the population fraction of one particular language, and $a$ the volatility characterizing the tendency to change state~\cite{abrams2003}. For high volatilities, $a<1$, coexistence of both languages has been found, whereas for $a>1$ one of the two languages becomes dominant~\cite{abrams2003}. Studying the AS model on networks, it has been shown that a decrease in network connectivity leads to a reduced parameter regime with language coexistence~\cite{vazquez2010agent}. In the same study, the AS model was also compared with a variation including a bilingual state, which has been observed to hinder coexistence.

Here, we propose a model which aims to capture the mechanisms of community formation and how these depend on the specific structure of the contact network. It describes the dynamics of a society in which individuals may  either be neutral (hold a default opinion) or belong to a minority with radical positions on a current topic. Examples of human behavior with such a dynamics are the membership and involvement in political parties or organizations, belonging to religious communities, the interest in leisure activities as online games, the behavior of consumers with new technologies, or the typical cycle of  addiction to chemicals. Neutral individuals may at some point in time make acquaintance with a new product or idea and subsequently become customers or adopt it as their default position. Hereafter, they can either like it and consolidate as their default state or not and return to neutral behavior. Additionally, if agents in a default state are in contact with fanatics (radical agents) 
they can also turn into radicals. Radical agents, in turn, may get bored or disappointed after a while reducing their level of commitment and turn back to a default position or even completely  leave the community and become neutral agents again.

There are two characteristic features of our model, defined mathematically in the next chapter, which are essential for its dynamics. First, we account for the fact that the population density may vary over the contact network, i.e. in addition to individuals with different states (opinions) we also consider empty sites. Second, the probability of a given individual to change its state is taken to be proportional to both the number of like and alike peers in its neighborhood. This product form of the transition rates is the main difference to previously studied non-linear voter-like models~ \cite{schweitzer2009non-linear,vazquez2008systems} and the AS model~\cite{abrams2003}. This form implies, that transition rates may change from a linear to a non-linear form upon changing the composition of the population. 

The outlined scheme bears similarities with the formation of mafias, in which normal citizens are the analog to default opinions, while mafias represent the radical minority. Neutral individuals are those who do not take active part in the society and that are represented by empty sites. Allegorically these sites may enter the community as citizens by  a \emph{birth event} and join the mafia later due to the social pressure of their vicinity. In this paper, for illustration we will use the metaphoric language of mafias.

We will show that the \emph{mafia model} yields a rich phase behavior which contains both coexistence of opinions and consensus on the default position in well-mixed societies. In scale-free networks, coexistence appears in a larger regime of the parameter space due to the effect of local interactions: the stability of the absorbing state (extinction) is lost. We will discuss the role played by the nature of interactions---\emph{one-to-all}, i.e. the whole neighborhood determines the state of single agents, as compared to one-to-one interactions like in the voter model---for the loss of stability of the absorbing state. Furthermore, we will show that  the inclusion of external fields in the model, representing the presence of elements  controlling the interactions (which we refer to as police individuals), gives rise to micro-phase separation where regions with different dynamics are observed to coexist in the system.   

This paper is organized as follows: We first describe the dynamics of the mafia model and then we analytically discuss the phase diagram expected for well-mixed populations. Then, we  analyze the behavior on structured societies showing different degrees of heterogeneity, which we  investigate through extensive stochastic simulations. We finally address the role of external control elements in the system and explore how specific distributions of those into highly or sparsely connected nodes modifies the system's behavior. An extended mean-field theory accounting for the local structure of networks is given in the Appendix.

\section{The mafia model}

\begin{figure}[t]
\centering
\resizebox{0.25\textwidth}{!}{%
\includegraphics{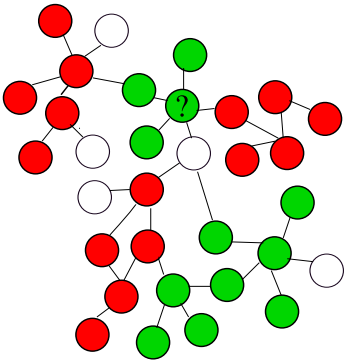}}
\caption{[color online] Individuals (red mafiosi, green citizens, white empty sites) are distributed in the nodes of a scale-free network. The composition of the neighborhood of the  agent who makes a decision, marked with a question mark, reads $(c, m,\phi) = (3/5,1/5,1/5)$.}
\label{fig: sfn cartoon}
\end{figure}

We model human interactions in social networks by graphs whose nodes allocate interacting agents and whose edges describe the relations between pairs of individuals (Fig.~\ref{fig: sfn cartoon}). The nodes of a network can either be empty ($\Phi$) or occupied by an individual. In the simplest case agents belong to one of two existing groups identified with what we refer to as two different \emph{strategies}, namely they either belong to the mafia ($M$) or to the group of lawful citizens ($C$). The network's size is assumed to be fixed and, therefore, the sum of the number of citizens, mafiosi, and empty places  (or their corresponding fractions) remains constant: $C+ M + \Phi =N$.

We employ an agent-based model in which the society evolves according to the following stochastic dynamics: (i) at empty nodes citizens are born with a site-independent fixed rate $b$, (ii) at occupied nodes individuals die with a constant rate $d$, and (iii) strategy changes take place in which citizens join the mafia at rate $w_{c\to m} (=: w_{cm})$ while mafiosi leave it at a rate $w_{m\to c} (=: w_{mc})$. Since individuals who reconsider their status, namely whether leaving or joining the mafia, are influenced by their surroundings, these rates are determined by the composition of the neighborhood of a given individual.
\begin{displaymath}
\centering
\xymatrix{
  &  {\Phi}\ar@<0.5ex>[dr]^{b}\\
  M\ar@<0.5ex>[rr]^{w_{mc}}\ar[ur]_{d}&& C\ar@<0.5ex>[ll]^{w_{cm}}\ar@<0.5ex>[ul]^{d} 
}
\end{displaymath}
As the neighborhood of a given agent we define the set of all its nearest neighbors, i.e.\ the set of all other agents connected directly to it through an edge of the network. In Fig.~\ref{fig: sfn cartoon}, for instance, the neighborhood of the citizen with the question mark contains three citizens (green), one mafioso (red), and one empty place (white). The composition of this neighborhood is characterized by the tuple $(c, m,\phi)$, where $c$, $m$ and $\phi$ denote the fraction of mafiosi, citizens and empty sites in the neighborhood of a given individual at a given instant of time. 

Single agents feel both the pressure from neighbors holding a different strategy and the support of neighbors who are  members of their peer group, motivating the following choice for the transition rates:
\begin{eqnarray}
w_{c\to m} = w_{cm} & = & m\, s_m \,(1-c),\label{eq: wcm}\\
w_{m\to c} = w_{mc} & = & c\, s_c\, (1-m)\label{eq: wmc}.
\end{eqnarray}
In this way, the pressure exerted on a given agent by individuals belonging to the adversary group is proportional to their fraction among the nearest neighbors, i.e.\ $m$ and $c$, respectively. The strength of this pressure, $s_m$ or $s_c$, which we also call  persuasiveness, is taken to be the same and constant for all individuals. At the same time, the supportiveness of the alike individuals attenuates this pressure through a multiplicative factor, $(1-c)$ and $(1-m)$, respectively. This product form of the transition rates implies that individuals do not reach a stable state but keep on changing their strategies unless the full neighborhood holds the same opinion and the agent is fully surrounded by alike agents. As a consequence the relative strength of the linear and non-linear term in the transition rates change with species frequencies, cf. $\omega_{cm}\sim m (1-c) =  m (m+\phi)$: If the abundance of the adversary species is much smaller than the fraction of empty places $(m \ll \phi$), the leading term to change an indiviuals' state is linear, whereas for species abundances significantly larger than the fraction of empty places the leading term is non-linear and the dynamics changes accordingly. This is distinct from previously studied non-linear opinion models like the Abrams-Strogatz model where the type of non-linearity remains fixed independent of the composition of the populations~\cite{abrams2003,vazquez2010agent}. In our notation the rates of the AS model read $\omega_{cm} = s_m m^a$ and $\omega_{cm} = s_c c^a$ with volatility $a$, and prestiges $s_{m/c}$ of ``languages''  $m$ and $c$.

Moreover, since the tuple $(c, m,\phi)$ denotes the overall composition of an individual's neighborhood, a given agent interacts with its neighbors in a \emph{one-to-all} scheme. As it will turn out, the ensuing dynamics and stationary states on networks are genuinely different from those obtained for a \emph{one-to-one} scheme where a given individual interacts pairwise with a single randomly chosen neighboring site. The main difference is that in \emph{one-to-all} schemes one has to account for a discrete set of possible neighborhood compositions. This feature invalidates standard mean-field theories even in models where one adds strong mixing. As shown in Appendix A, an extended mean-field theory accounting for these discrete set of possible neighborhood compositions explains why a one-to-all interaction in structured societies induces changes in the phase diagram.

In this paper, we will focus on the \emph{asymmetric case} in which mafiosi are more persuasive (much stronger) than citizens, $s_m \gg s_c \approx 0$. For this case, the transition rate $w_{mc}$ vanishes and individuals can leave the mafia only rather indirectly, namely via a death-birth process. The more general case yields less interesting dynamics;  a full discussion may be found in \cite{balbas2010diss}. 

Decision processes may, in general, also be influenced by many external factors, such as mass media, independent of the interacting actors. More particularly, societies may provide means to regulate their conflicts and protect  citizens against the damage of mafias. One would like to account thus for the role of control elements such as police in actual societies. Control elements do not participate directly in the social dynamics by interacting with the agents and evolving in time, but specifically control the relations between pairs of individuals. They are located at the edges connecting nodes between two individuals of a social network. The fraction of total edges allocating control elements is $p$. The model includes their two-fold catalytic role activating the transition $m\to c$ and  inhibiting the transition $c\to m$: 
\begin{eqnarray}
w_{cm} & = & m s_m (1-c) (1-p),\\
w_{mc} & = & (c s_c + s_p p)(1-m).
\end{eqnarray}
The control elements protect citizens from the mafia's pressure, attenuating their strength proportional to the factor $(1-p)$. Simultaneously they persuade individuals to leave the mafia proportional to their fraction $p$ and strength $s_p$. The persuasiveness adds to that of the citizens, both being still attenuated by the presence of mafiosi as before.

\section{Mean-field approximation}

For well-mixed populations in which every individual interacts with all other agents in the system, the problem we pursue to solve is topologically equivalent to having the agents allocated on the nodes of a complete graph. Then the \emph{local field} which every agent experiences is the same as the global field, given by the average relative abundance in the whole population. This simplification leads to a mean-field approach for well-mixed societies which allows an analytical characterization of the problem. We note that this type of mean-field limit does not correspond to the limit of fast diffusion on networks if the interaction is \emph{one-to-all} as is the case in our mafia model. Then, the dynamics of the network-structured society is actually better described within an extended mean-field theory which accounts for the fact that for each given node there is a finite set of neighborhood compositions, cf. Appendix A. In the following we discuss the standard mean-field approach in order to highlight the novel effects introduced by the network structure and the 
 \emph{one-to-all} interaction scheme.
 
We non-dimensionalize by choosing the inverse death rate, $d^{-1}$, as our basic time scale and define dimensionless time and all other parameters accordingly: $\tau = td$, $\beta = b/d$, $\sigma_i = s_i/d$, $\omega_{ij} = w_{ij}/d$, with $i,j \in \{c,m\}$. Then, the mean-field equations for the time evolution of the different population fractions read:
\begin{eqnarray}
\dot{c}  & = & \phi\,\beta + m\,\omega_{mc}  - c\,\omega_{cm}  - c,\label{eq: dotc}\\
\dot{m} & = &  -m\,\omega_{mc}  + c\,\omega_{cm}  - m,\label{eq: dotm}\\
\dot{\phi} & = & -\phi\,\beta  +  (1-\phi ),\label{eq: dotphi}
\end{eqnarray}
where a dot signifies a time derivative with respect to $\tau$. From Eq.~\eqref{eq: dotphi} we immediately see that the stationary fraction of empty places is independent of the social dynamics encoded in the transition rates $\omega_{cm}$ and $\omega_{mc}$, but only depends on the dimensionless birth rate $\beta$:
$$\phi=\frac{1}{1+\beta}\,.$$
In other words, the birth rates determines how densely populated the system is. Inserting this result together with the constraint $c+m+\phi=1$ into Eq.~\eqref{eq: dotm} yields an implicit equation, whose solutions are the fixed points or stationary states of the population dynamics:
$$ 
m\,\omega_{mc}(c,m,p) - \left(\frac{\beta}{1+\beta} - m\right)\,\omega_{cm}(c,m,p)  + m = 0.
\label{eq:master eq}
$$

In the following, as noted above, we focus on the asymmetric case where mafiosi are much stronger than citizens, $\sigma_m \gg \sigma_c\approx 0$, and use the simplified notation $\sigma:= \sigma_m$. Then, in the absence of control elements, the mean-field equations reduce to:
\begin{eqnarray}
\dot{c}  & = & \phi\,\beta -\sigma\,m\,(1-c)\,c -c,\label{eq: dotc fam}\\
\dot{m} & = &  \sigma\, m\,(1-c)\,c  - m,\label{eq: dotm fam}
\end{eqnarray}
whose fixed points are:
\begin{equation}
m^0=0, \qquad m^{\pm} = \frac{1}{2}\left(\frac{2\beta}{1+\beta}-1\pm \sqrt{1-4 / \sigma}\right).\end{equation}

\begin{figure}[h]
\centering
\resizebox{0.4\textwidth}{!}{%
  \includegraphics{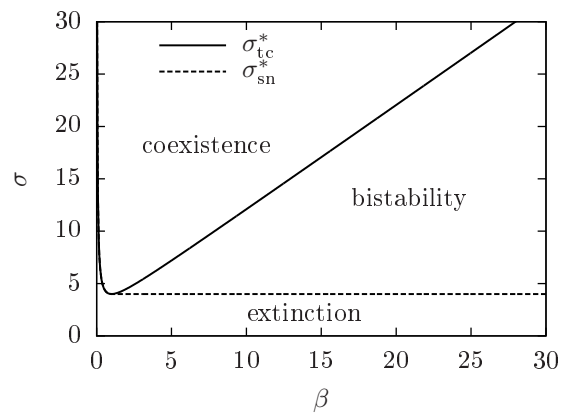}}
\caption{Phase diagram as a function of the birth rate $\beta$ and persuasiveness $\sigma$. Three regions with different regimes (coexistence, bistability, extinction) are shown, separated by a transcritical (solid) and a saddle-node (dashed) bifurcation.}
\label{fig: stability diagram} 
\end{figure}

The ensuing phase diagram (Fig.~\ref{fig: stability diagram}) shows three
different regimes: coexistence, bistability, and mafia extinction.
For a given birth rate $\beta$, mafiosi unavoidably get extinct below the threshold strength
$\sigma^*_{\text{sn}}$, while for  strengths larger than $\sigma^*_{\text{tc}}$
coexistence is the only stable solution. In the intermediate
regime, $\sigma^*_{\text{sn}} < \sigma <\sigma^*_{\text{tc}}$, the
system is bistable and the stationary solution critically depends on
the initial conditions. It turns out  that the minimal strength of
mafiosi to ensure that they do not get extinct increases with the
birth rate of citizens, although very weak mafiosi die out
independently of the birth rate. 

\begin{figure}[h]
\centering
\resizebox{0.4\textwidth}{!}{%
\includegraphics{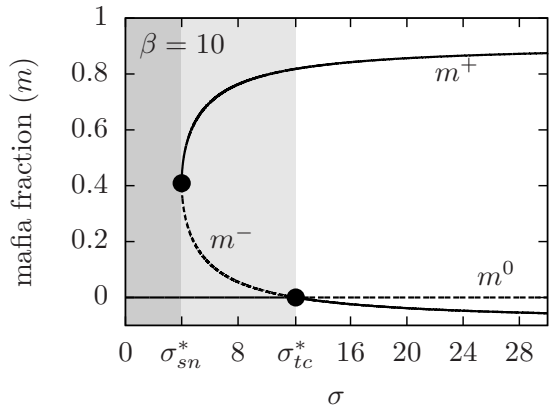}
}
\caption{Bifurcation diagram for a fixed birth rate $\beta=10$ as a
  function of the strength parameter $\sigma$. The solid and dashed
  lines represent stable and unstable solutions, respectively. The two
  solid circles stand for the saddle-node bifurcation, at which the solutions
  $m^{\pm}$ become complex, and the transcritical bifurcation, above which the 
  absorbing state $m^0$ becomes unstable, respectively. In the dark grey and white areas the
  system is monostable---extinction and coexistence are the stable
  states respectively---and bistable in the light grey region.}
\label{fig: bd}
\end{figure}

For a  particular value of the birth rate $\beta\neq 0$ (Fig.~\ref{fig: bd}), the absorbing state $m^0$ becomes unstable via a transcritical bifurcation for a persuasiveness larger than the threshold value $\sigma^*_{\text{tc}} = (1+\beta)^2/\beta$, while the coexistence state $m^+$ is real only above the saddle-node bifurcation at $\sigma^*_{\text{sn}}=4$ (for $\beta >1$). In the regime between these two values the system is bistable. The third solution $m^-$ is unstable for all meaningful values.  The stationary fraction of mafiosi in the coexistence state increases with their strength up to the asymptotic value $m^+ =\rho = \beta/(1+\beta)$ in the limit of infinite strength, $\sigma \to \infty$, where $\rho$ is the fraction of populated nodes.  

\section{Structured social networks}

Individuals in actual societies do not interact with the whole population, but rather only with agents in their immediate vicinity. A given agent who updates her strategy is influenced by her nearest neighbors, so that the fraction of citizens, mafiosi, and control elements entering the transition rates \eqref{eq: wcm} and \eqref{eq: wmc} are now given by the instantaneous composition of her neighborhood. The population frequencies $(c,m, \phi)$  are restricted to a \emph{finite set} of possible combinations determined by the size of the neighborhood. They are thus no longer continuous as in the mean-field case discussed in the previous section, but take a \emph{discrete} set of values determined by the composition of the neighborhood of a given node.

We investigate the behavior of the mafia model on scale-free networks (SFN) whose degree distribution scales as a power law $p(k)\propto k^{-\gamma}$. The parameter $\gamma$ is related to the heterogeneity of the network and increases for more  homogeneous structures. Scale-free networks~\cite{albert2002statistical,cohen2003scale} exhibit very interesting features suitable for the modelling of social systems, such as  large heterogeneity---with a non-negligible fraction of highly connected nodes---and relative small average path-lengths, i.e.\ the average over the shortest distance between all pairs of nodes. The latter scales logarithmically with the network's size, as it has been observed in human communities~\cite{albert2002statistical}.

The scale-free networks used in this work have been generated following 
the uncorrelated model \cite{catanzaro2005generation}, for which the resulting average degree $\langle k \rangle$ is a function of $\gamma$. This method imposes a maximal cutoff for the degree, automatically yielding the average degree for a given network size; see \cite{catanzaro2005generation} for further details. In this work, we have used networks with increasing $\gamma$ ranging from $2$ to $4$, and corresponding decreasing average connectivities $\langle k \rangle$ varying from $7.6$ to $2.4$.

We have performed stochastic simulations employing an agent-based model following the dynamics described above. Individuals belonging to both populations were taken as randomly distributed with specific initial fractions $m_0$ and $c_0$. We used random sequential updating according
to the dynamical rules defined previously. The results discussed in this paper were averaged over $1000$ runs. Starting from a given initial state, the system evolved until it reached a quasi-stationary state; since the system has an absorbing state, $m^0=0$, it will be unavoidably reached if one waits long enough. The population remained in this state for a long enough time window where the relevant observables were measured. We have carried out a finite size analysis and assessed that the systems reaches a quasi-stationary state at a time $\tau '$ proportional to the system's size, $\tau '= \lambda N$, with the proportionality factor $\lambda$ ranging from $0.01$ to $0.015$ (not shown). 

\begin{figure}[b]
\centering
\subfigure[complete graph]{%\resizebox{0.24\textwidth}{!}{%
\includegraphics{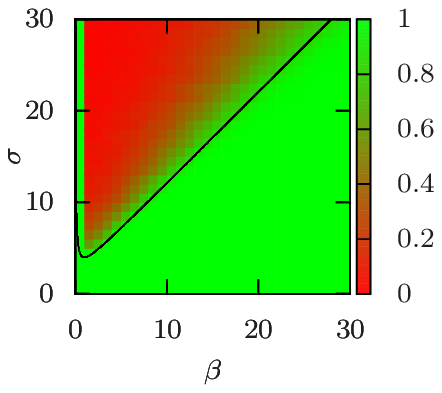}}%}
%\hspace{1ex}
\subfigure[$\gamma=3$ SFN]{%\resizebox{0.24\textwidth}{!}{%
\includegraphics{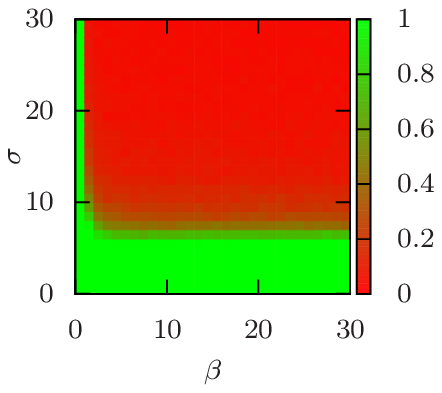}}%}
\caption{[color online] Extinction probability---as encoded in the side bar---for initial conditions with a very small fraction of mafiosi, $m_0=5\cdot 10^{-4}$, for (a) a complete graph of size $N = 6000$  and (b) a scale-free network (SFN) with $\gamma=3$ and $N=8000$ sites ($\langle k \rangle=3.16$). The solid black line represents the deterministic mean-field prediction separating a region of coexistence above it from a regime in which mafiosi get extinct below it.
}
\label{fig: FAM p_ext}
\end{figure}

We characterize the stationary state of the system in terms of the extinction probability $p_{\text{ext}}$, which is the probability that mafiosi got extinct at some time $\tau '$ within the quasi-stationary time window. The ensuing phase diagrams, as shown in  Fig.~\ref{fig: FAM p_ext}, are qualitatively different for well-mixed graphs and scale-free networks. In particular, the topology of the underlying network strongly matters in the parameter regime where mean-field theory predicts bistability, i.e. a dependence of the stationary state on the initial state of the population.
We have specifically analyzed the limiting case in which the initial population of mafiosi vanishes $(m_0\to 0)$. In this limit, the standard mean-field theory asserts that mafiosi go extinct for all of the bistable area. This is indeed the case for well-mixed societies (complete graphs), where the results of our stochastic simulations closely resemble the prediction of the (deterministic) mean-field theory. There are some minor differences in the actual position of the phase boundary which we attribute to stochastic effects; see Fig.~\ref{fig:  FAM p_ext}a. In contrast, coexistence is systematically found in the bistable regime (of mean-field theory) for all studied networks (scale-free networks with $\gamma$ ranging from $2$ to $4$)\footnote{We have also performed simulations on square grids which show similar behavior as scale-free networks~\cite{balbas2010diss}.}. An exemplary phase diagram for a network with $\gamma = 3$ and a small initial mafiosi fraction is shown in Fig.~\ref{fig:  FAM p_ext}b. 

We attribute this interesting feature of what one could call an ``evolutionary stability'' of mafiosi to the locality of interactions on networks or regular graphs. In structured societies, the one-to-all character of interactions becomes relevant, as averaged frequencies suitably reproduce interactions between two individuals (one-to-one) but fail to describe interactions with a whole vicinity. An extended mean-field analysis, accounting for the discrete set of possible neighborhood compositions is thoroughly discussed in Appendix A. This extended mean-field theory gives a phase diagram where the bistable regime is drastically reduced in favour of the coexistence regime, in accordance with the above numerical results.

What is the micro-dynamics giving rise to these results?  A vanishing small fraction of mafiosi gives rise to negligible transition rates $\omega_{cm}$ within a standard mean-field approach and consequently the mafia dies out in a short period of time. However, if interactions are local, citizens' vicinities containing any mafioso yield considerable transition rates for updating citizens to become mafiosi. Under which circumstances can a few isolated mafiosi invade a citizen population? The following discussion gives a set of heuristic arguments which are not meant to be quantitative but illustrate the basic mechanism behind the observed evolutionary stability. 

\begin{figure}[h]
\centering
\subfigure[isolated]{\resizebox{0.2\textwidth}{!}{%
\includegraphics{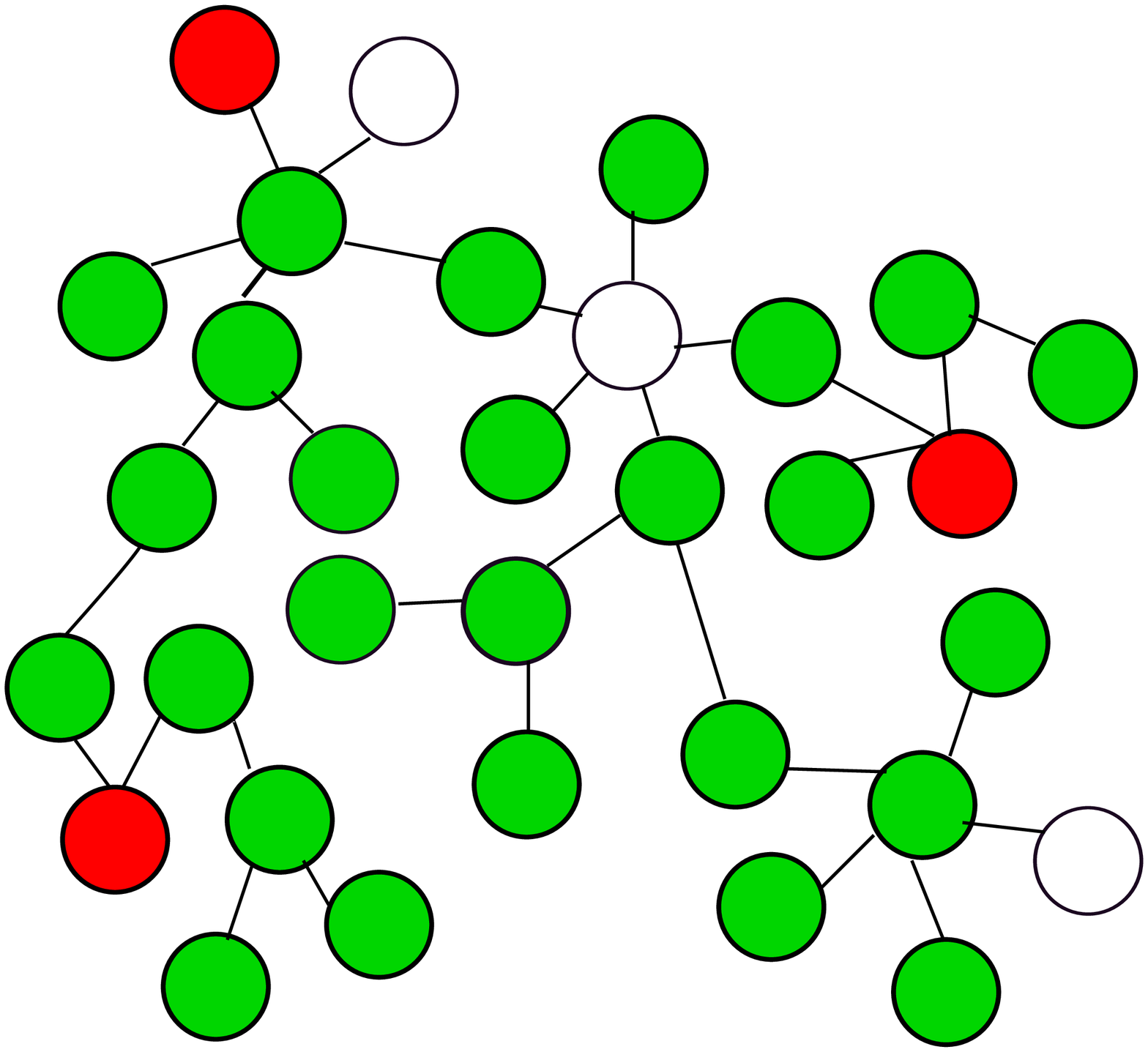}}}
\subfigure[invasion]{\resizebox{0.2\textwidth}{!}{%
\includegraphics{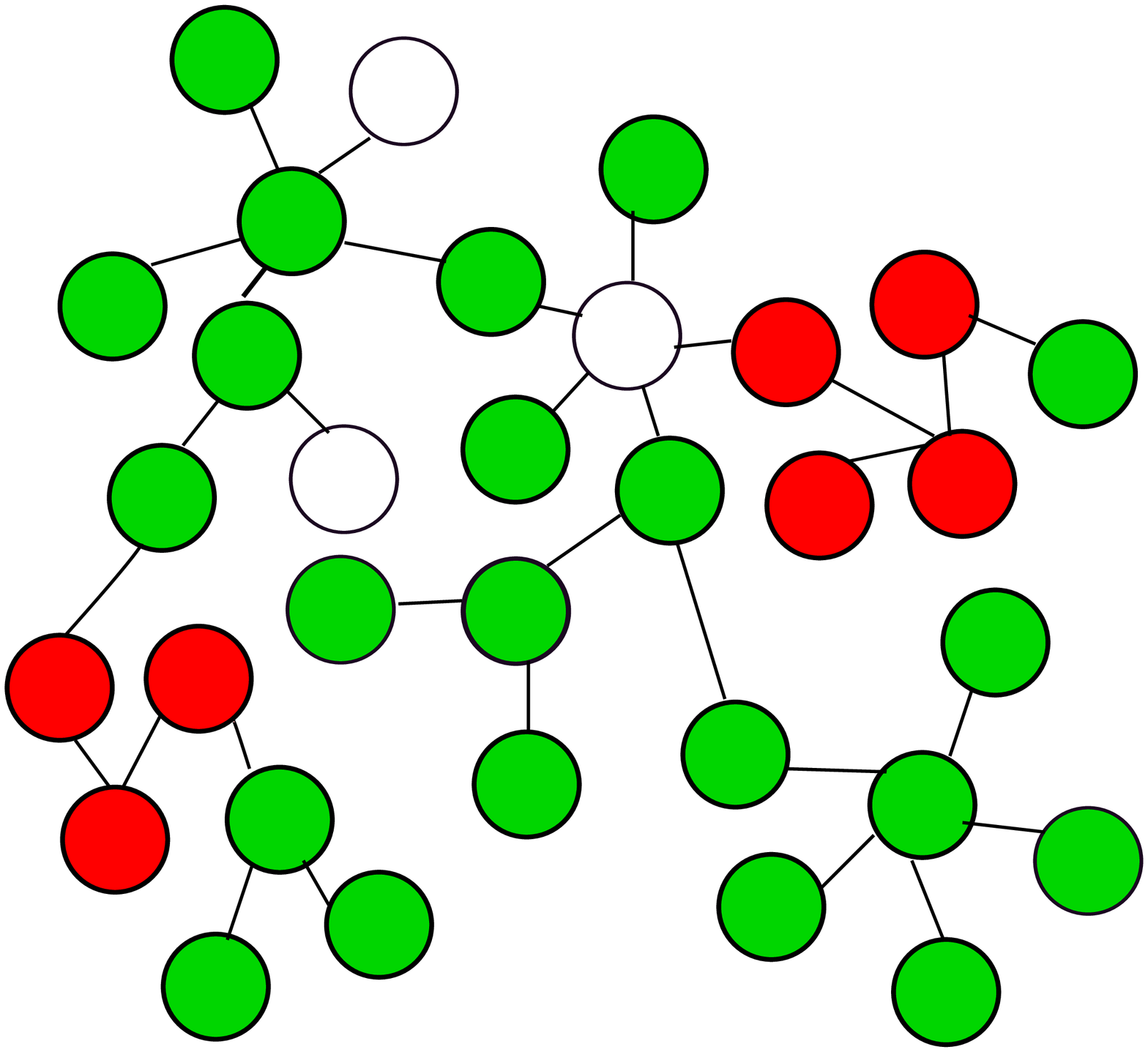}}}\\
\subfigure[expansion]{\resizebox{0.2\textwidth}{!}{%
\includegraphics{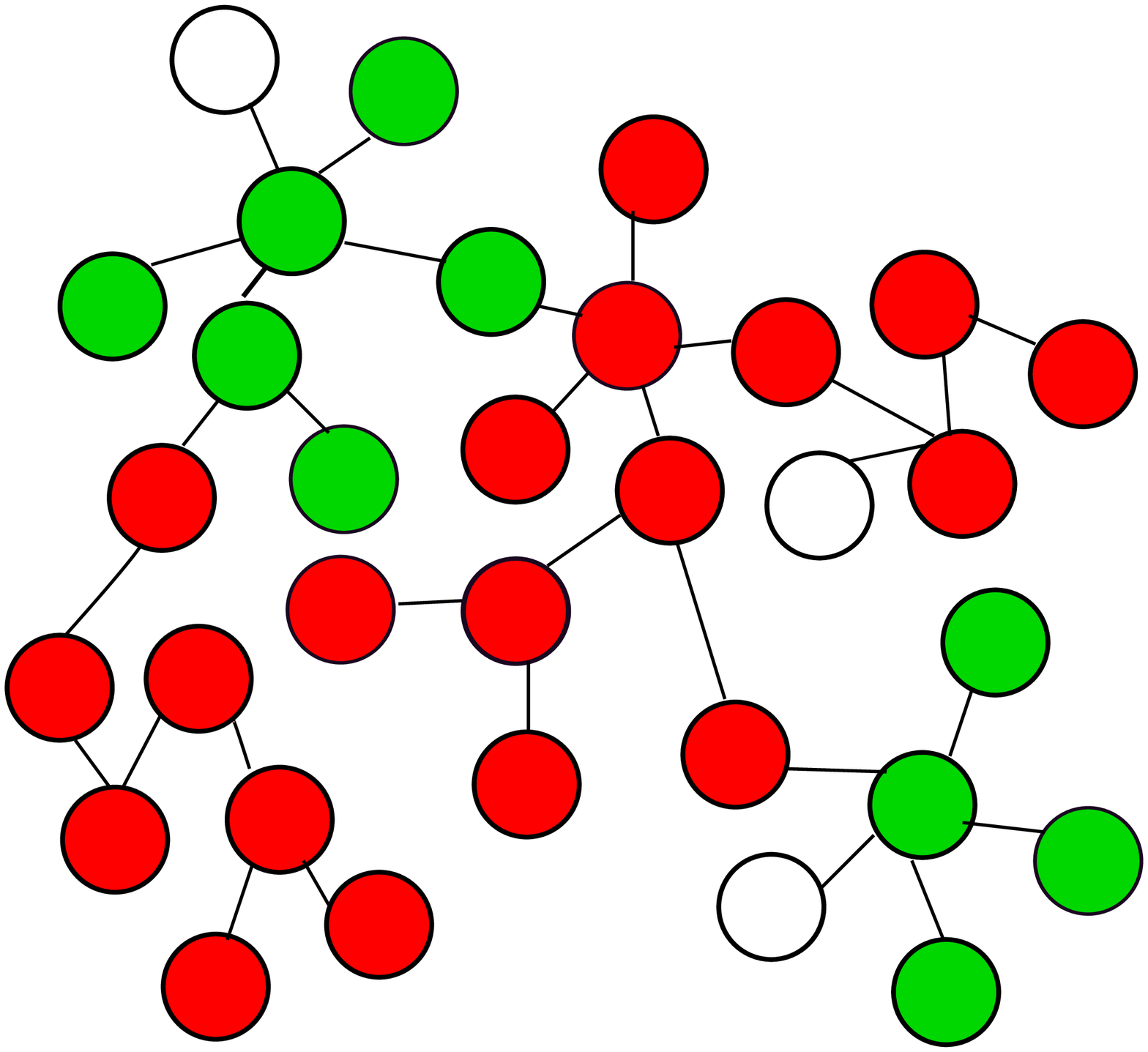}}}
\subfigure[saturation]{\resizebox{0.2\textwidth}{!}{%
\includegraphics{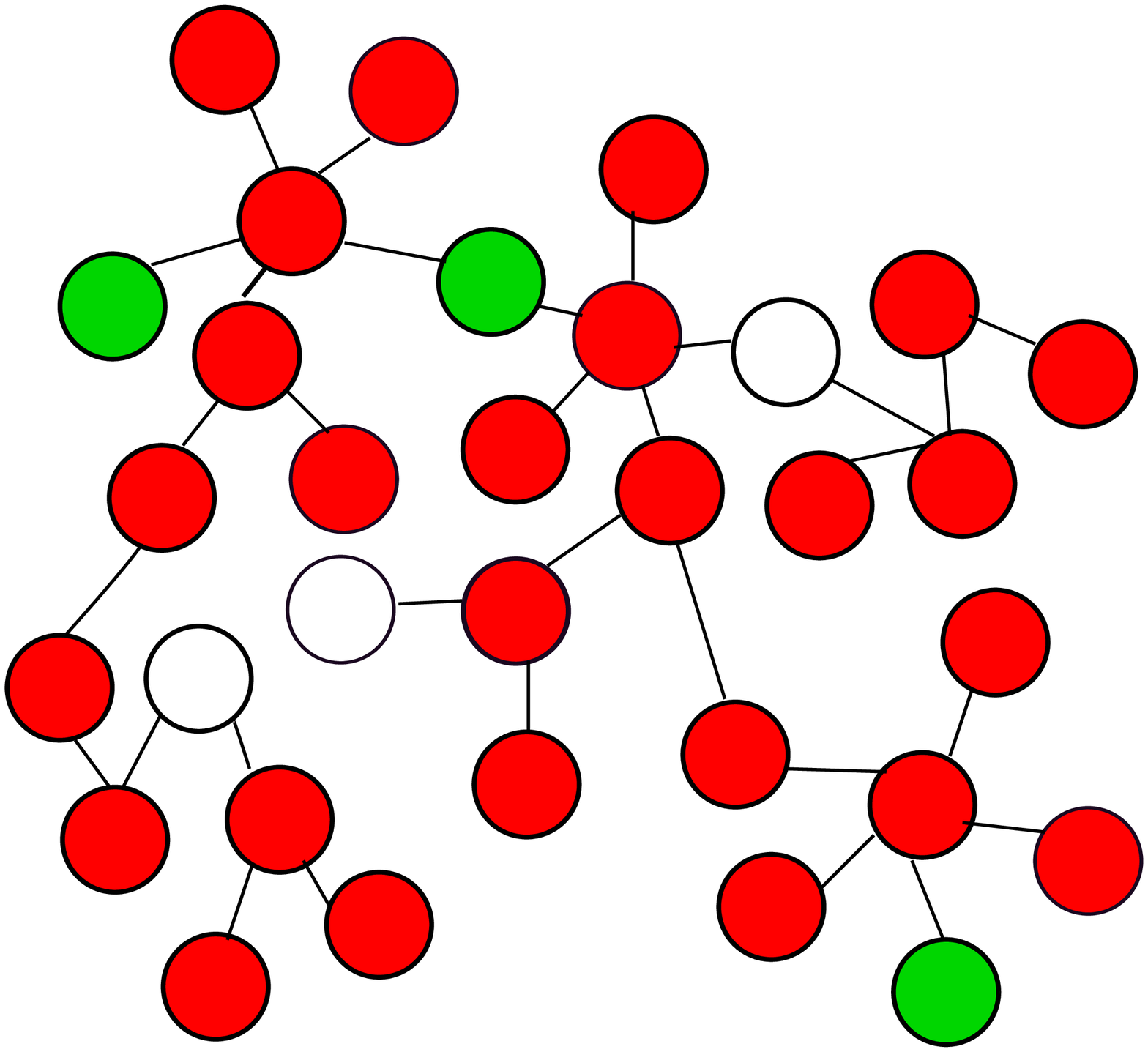}}}
\caption{[color online] Invasion-expansion-saturation
  process. \emph{Invasion}: A few isolated mafiosi invade their nearest
  neighbors instead of dying, which leads to the formation of very small clusters of mafiosi.  \emph{Expansion}: these small clusters expand
 as more and more citizens successively join the mafia.  \emph{Saturation}: the
  society reaches a stationary state with a dynamic balance between birth and death processes.}
\label{fig: invasion-expansion-saturation}
\end{figure}

Starting from some isolated mafiosi, the dynamics may be separated into three phases as illustrated in Fig.~\ref{fig: invasion-expansion-saturation}: invasion-expansion-saturation.
First, the isolated mafiosi must \emph{invade} some neighboring cell before dying, i.e.\ the probability for any citizen in the mafioso's neighborhood to change her strategy must be larger than her death probability. The probability for a mafioso at site $i$ to invade any neighboring cell is the sum over the probabilities for all her $k_i$ neighbors (assumed to be citizens) to change strategy, i.e.\ $p_{\text{inv}}=\sum_{j=0}^{k_i} p^j_{c\to m} =\sum_{j=0}^{k_i} \sigma m (1-c)\Delta\tau$, which yields  $k_i \sigma (1/k_j)^2 \Delta\tau$ considering that every citizen has a single mafioso in her vicinity. If the structure is homogeneous enough, so that one may consider  similar  connectivities, i.e.\ $k_i\sim k_j$,\footnote{Note that this approximation is formally valid only for SFN with large $\gamma$ or random graphs.} then the invasion probability becomes $ \sigma / k_i\,\Delta\tau$. The invasion probability is larger than the death probability for mafia's strength  larger than the degree $\sigma_{\text{tinv}}> k_i$. This heuristic argument nicely explains the results obtained from simulations which show that the extinction probability decreases with the network's average degree $\left< k \right>$ (not shown). Say now the mafioso has not died out but managed to invade some neighboring cells and thereby created  small mafia clusters. In order for these clusters to survive they have to \emph{expand} as groups to achieve a stable coexistence state in the population. Mafia clusters will grow if the fraction of citizens becoming mafiosi at their interfaces is larger than the fraction of those citizens gained via birth process. Assuming similar fractions of citizens and mafiosi in the vicinity of an agent located at the  interface, $m \sim c \sim 1/2$, and large enough birth rates\footnote{This assumption implies that one can neglect empty sites and approximate the fraction of citizens and mafiosi at both sides of the cluster boundary by $1$.  The approximation becomes better with an increasing size of the vicinities. For small neighborhoods, taking empty sites into account would make it easier for the mafiosi to expand their territory as citizens have less peer protection. Therefore, accounting for empty sites would slightly lower the threshold strength such that our approximation can be regarded as an upper bound.} $\beta$, the fraction of born citizens is given by $l_{\text{int}} \phi \beta$ and those becoming mafiosi $l_{\text{int}} \omega_{cm}\sim l_{\text{int}} \sigma m (1-c) \sim l_{\text{int}} \sigma /4$, where $l_{\text{int}}$ is the length of the cluster interface. Mafia clusters will thus grow if their strength becomes larger than some threshold value, $\sigma > \sigma_{\text{texp}} =4$. Finally, mafia clusters will expand to the point in which there are no citizens left at which expense they can grow. This \emph{saturation} takes places for a residual fraction of citizens, which  equals the fraction of born citizens at empty places. A dynamic stationary state has thus been reached at which both populations coexist due to the asymmetric birth of citizens---otherwise citizens would get extinct. These heuristic arguments for the growth process in structures, supported by the results of the stochastic simulations, nicely evidences that above some threshold value for the mafia's strength, the mafia can always invade a citizens population. This threshold is given by the larger of the following two values: the mafia strength required to invade neighboring sites, $\sigma_{\text{tinv}}$, or for small clusters to expand, $\sigma_{\text{texp}}$, i.e.  $\sigma > \mathrm{max}(\sigma_{\text{tinv}}, \sigma_{\text{texp}}) = \mathrm{max}(\langle k \rangle, 4)$. 

In the above line of reasoning, we have assumed that very node has the same connectivity, which is true only for random networks. However, since mafia's expansion is a process which depends on the \emph{local} dynamics, we suppose that our heuristic approach should remain valid for scale-free networks. Moreover, the existence of regions with a connectivity significantly smaller than the average in scale-free networks suggests that coexistence is more likely to be found there than for the corresponding random graph with the same average degree.

Another interesting result is that in the limit of complete random graphs, $\langle k \rangle\to N$, the above heuristic approach gives a threshold strength $\sigma \geq 1+\beta$ for mafiosi to survive, which agrees well with the standard mean-field analysis---see Fig.~\ref{fig: FAM p_ext}a.

\section{The role of control elements}

Societies can regulate themselves by introducing control elements which fight the expansion of mafias. Here, we consider control elements (police) which are randomly attached to the edges of a given graph, such that a fraction $p$ of the total number of edges---and thus of pairs of individual relations---are controlled. With an increasing number of police in the asymmetric model the mafiosi also need to enhance their strength/persuasiveness in order to survive. For simplicity, we consider parameters such that $\sigma_m=\beta = \sigma$ corresponding to the diagonal in Fig.~\ref{fig: stability diagram} where only two regimes are
observed in the mean-field limit, namely extinction and bistability. 

\begin{figure}[t]
\centering
\subfigure[complete graph]{%\resizebox{0.24\textwidth}{!}{%
\includegraphics{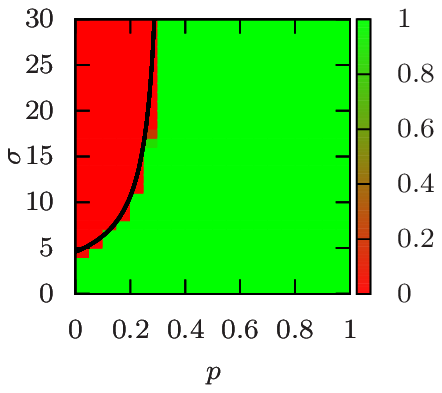}}%}
\subfigure[SFN $\gamma=4$]{%\resizebox{0.24\textwidth}{!}{%
\includegraphics{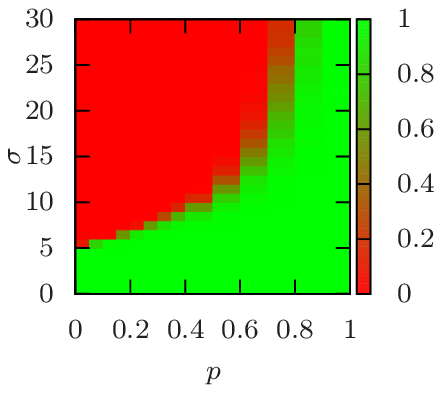}}%}
\caption{[color online]  Extinction probability for a policed system on (a) a complete graph (well-mixed population) with $N=6000$ and on (b) the nodes of a scale-free network with $\gamma=4$ ($\langle k \rangle=2.45$) and $N=8000$. The solid line in (a) represents the separatrix predicted by the deterministic mean-field theory for the given initial conditions $m_0=1/2$.}
\label{fig:policed_structures}
\end{figure}

A standard mean-field analysis -- equivalent to a well-mixed population or alternatively a complete graph -- including control elements with strength $\sigma_p = \sigma$ shows the same type of regimes as for the model without police: For small police fractions the system is bistable and coexistence or extinction is obtained depending on the initial composition of the population. Above a certain police fraction, the mean-field analysis predicts extinction of mafiosi as the only stable fixed point; see Fig.~\ref{fig:policed_structures}a for a phase diagram with initial condition $m_0  = 0.5$.  The mafia's strength required to achieve coexistence is increased with the presence of control elements. 

\begin{figure}[b]
\centering
\resizebox{0.45\columnwidth}{!}{%
\includegraphics{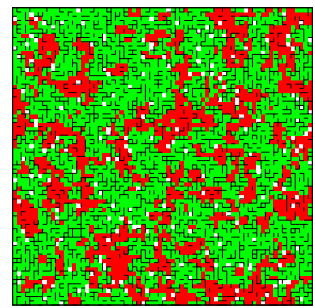}}
\caption{[color online] Snapshot of the population structure on a lattice with a police fraction $p=0.3$ and size $N=4900$. Red cells are occupied with mafiosi, green with citizens, and white cells correspond to empty nodes. The black segments represent policed connections between adjacent cells. Citizens and mafiosi cluster preferentially around policed and unpoliced areas, respectively.}
\label{fig: snapshot policed lattice}
\end{figure}

On scale-free networks, the local character of interactions again leads to an enlarged coexistence region as compared to the well-mixed case, cf. Fig.~\ref{fig:policed_structures}b. However, there is a significant difference to networks without control elements. In the present case, the coexistence regime is enlarged beyond the bistable regime of mean-field theory and now also covers an area in parameter space where the system was monostable  for well-mixed populations (i.e.\ where extinction was the only stable state); cf. Fig.~\ref{fig:policed_structures}a and b.  The reason for this anomalous behavior is that the control elements introduce a kind of \emph{quenched disorder}, which from thermodynamic systems is known to strongly affect phase behavior.  Indeed, we have checked by numerical simulations (data not shown) that coexistence is found only in a parameter regime where mean-field theory predicts bistability, if the control elements are allowed to diffuse on the network (annealed disorder). Analyzing the population dynamics on a local scale, we observe that the presence of control elements leads to a kind of micro-phase separation in the society: while citizens preferentially populate policed regions, mafiosi aggregate in unpoliced areas. The unpoliced areas display the same dynamics as the fully asymmetric model, in which species coexist for mafias' strength larger than some threshold, $\sigma > \sigma_t$. Fig.~\ref{fig: snapshot policed lattice} shows a snapshot of a typical population structure on a square lattice, where both regimes are clearly recognizable.

\begin{figure}[h]
\centering
\resizebox{0.4\textwidth}{!}{%
\includegraphics{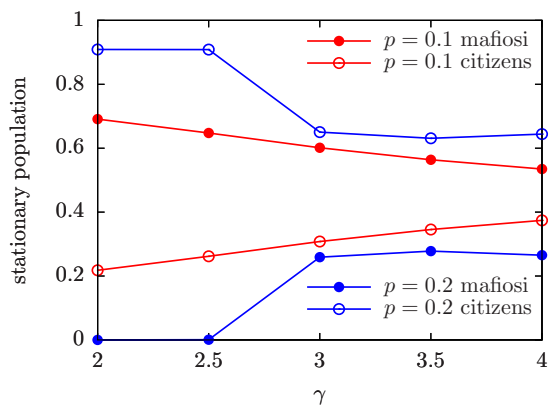}
}
\caption{[color online] Stationary fractions of mafiosi and citizens as a function of the network homogeneity $\gamma$ for two different values of the police fraction $p$. The values of $p$ were chosen such that for $\gamma = 2$ the stationary state corresponds either to coexistence ($p=0.1$, red), or mafia extinction ($p=0.2$, blue). Hence, for $p=0.1$ citizens are the minority species, while for $p=0.2$ mafiosi are the minority. Solid and open symbols represent mafia and citizens fractions, respectively. In both cases the respective minority species increases in population size with increasing network homogeneity $\gamma$, or, equivalently, decreasing average degree $\langle k \rangle$. The average degrees with increasing $\gamma$ ticks read $\langle k \rangle= 7.63, 4.31, 3.16, 2.69, 2.45$. Data are obtained for the following parameters: $N=10 000$, $m_0=c_0=1/2$, $\sigma=\beta=10$. } 
\label{fig: stationary population policed gamma}
\end{figure}

Micro-phase separation results in an increase of the respective minority's population size with increasing network homogeneity $\gamma$, or, equivalently, decreasing average degree $\langle k \rangle$, cf. Fig.~\ref{fig: stationary population policed gamma}. This can be understood as follows: Increasing network homogeneity implies smaller neighborhood sizes and larger average paths between pairs of nodes, and thereby increasingly hinders mixing of mafiosi and citizens. As a consequence, phase separation into isolated mafia and citizen clusters becomes more pronounced. Then, the respective minority species can find niches where they can grow to population sizes larger than in a corresponding well-mixed environment. 

\begin{figure}[h]
\centering
\resizebox{0.4\textwidth}{!}{%
\includegraphics{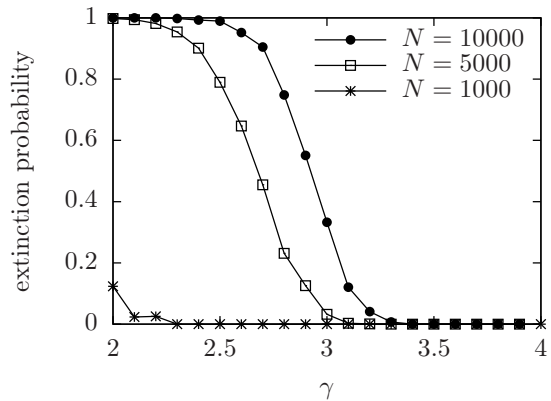}
}
\caption{Extinction probability of mafiosi for a fixed police fraction $p=0.3$ as a function of network homogeneity $\gamma$. With increasing overall population size $N$ the transition between certain extinction and certain survival of mafiosi becomes very pronounced indicating that there might be a phase transition as a function of $\gamma$. Data were obtained for the following set of parameters: $\sigma = \beta =10$ and $m_0=c_0=1/2$. The average degrees with increasing $\gamma$ ticks read $\langle k \rangle= 7.63, 4.31, 3.16, 2.69, 2.45$.}
\label{fig: extinction policed gamma}
\end{figure}

A finite size analysis of the extinction probability  with the increasing heterogeneity of graphs, for a fixed police fraction ($p=0.3$), is shown in Fig.~\ref{fig: extinction policed gamma}. The increasing slope of the extinction probability with $N$ suggests that as $N \to \infty$ there might be a phase transition separating an extinction from a coexistence regime. To make this point conclusive, however, would require to simulate even larger system sizes. Our main point is that for sufficiently  homogeneous networks (large $\gamma$) the phase separation between policed and unpoliced areas allows survival of mafiosi in the unpoliced areas for police fractions which otherwise would suffice for their extinction in
well-mixed and heterogeneous populations. 

\subsection*{Targeted police distribution}

In the same way that vaccination during an epidemic is more effective if targeted to highly connected nodes or hubs \cite{pastor2002immunization,anderson1982directly}, one might suppose that the distribution of control elements also matters in eradication of mafias.
To explore this question we investigate the effect of two specific distributions $q_\alpha (e_{ij})$ of control elements:
\begin{eqnarray}
q_1(e_{ij}) & =&  \frac{d_i d_j}{\sum_{ij}d_i d_j},\\
q_2(e_{ij}) & = & \frac{(d_i   d_j)^{-1}}{\sum_{ij}(d_i   d_j)^{-1}}.
\label{eq: police probability distribution}
\end{eqnarray}
These probability distributions for a control element to be allocated  at an edge $e_{ij}$ connecting nodes $i$ and $j$ are taken as either directly ($q_1$) or inversely ($q_2$) proportional to the product of the degrees of the nodes connected by the edge, $d_id_j$.  Therefore, the first distribution favours police attachment close to highly connected nodes, whereas the second promotes control elements to surround sparsely connected nodes. 

The network topology  should be the key element for understanding the relevance of a specific police distribution. Following the analogy with the problem of vaccination, one would
expect that surrounding network hubs with control elements should hinder the proliferation of mafias, since individuals in hubs are expected to be more influential. Although highly connected individuals influence the decision process of a larger number of agents, their influence on the targeted individual is as important as that of sparsely connected agents.
Therefore, controlling the population in hubs is not decisive for the system's dynamics. Actually, due to the concentration of control elements around hubs, the fact that a large number of small nodes are left unpoliced plays a more important role in the time evolution of the system than blocking the hubs.  A large number of not so well connected agents are free to influence the  society and drive it to a stationary state in which mafias do not get extinct. 

\begin{figure}[h]
\centering
\resizebox{0.4\textwidth}{!}{%
\includegraphics{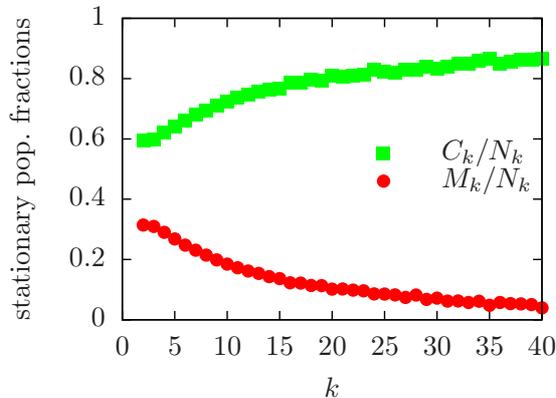}}
\caption{[color online] Stationary fraction of nodes of degree $k$ populated by citizens (free squares) and mafiosi (red circles) for the case where control elements preferentially attach to hubs ($q_1$). The simulations were run for $\sigma =10$, $p=0.2$ in a $\gamma =2.5$ scale-free network of size $N=8000$.}
\label{fig: population deg-dist}
\end{figure}

The micro-phases induced by the presence of control elements are
now localized around  highly and sparsely connected nodes depending
on the police distribution. We have observed 
that policed areas are quickly occupied by citizens during the initial stages of the 
dynamics, while at later times the dynamics evolves in the unpoliced regions. In the stationary state, there is also a clear division in the degree of the nodes occupied by citizens
and mafiosi: the population distribution follows the police distribution with citizens clustering around policed edges;  see Fig.~\ref{fig: population deg-dist} for an example with control
elements drawn from the $q_1$-distribution.

\begin{figure}[h]
\centering
\resizebox{0.4\textwidth}{!}{%
\includegraphics{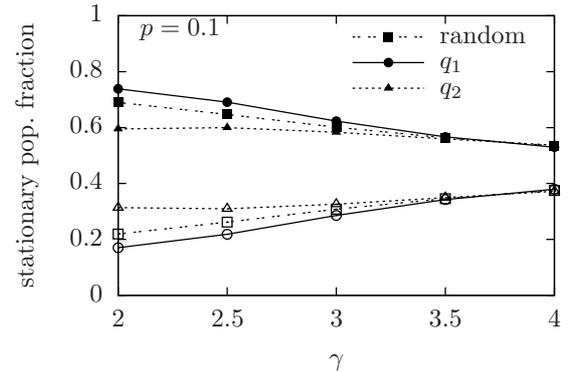}}
\caption{Stationary population fractions for different distributions
  of the control elements and fixed police fraction $p=0.1$ as a
  function of the increasing homogeneity of the scale-free network ($\gamma$). The solid
  symbols represent the mafia fraction, the empty ones the citizen
  population in the stationary state.  The average degrees with increasing $
\gamma$ ticks read $\langle k \rangle= 7.63, 4.31, 3.16, 2.69, 3.45$.}
\label{fig: police distributions}
\end{figure}

\begin{figure}[h]
\centering
\resizebox{0.4\textwidth}{!}{%
\includegraphics{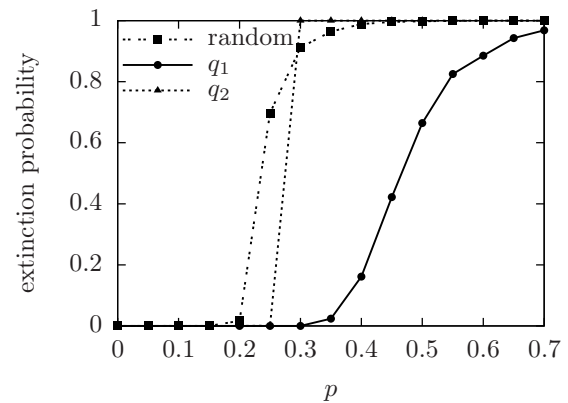}}
\caption{The police fraction needed to achieve extinction of mafiosi
  on a $\gamma=2.5$ ($\langle k \rangle = 4.31$) scale-free network of size $N=10000$ crucially
  depends on the distribution of control elements on the network. A
  larger mafia fraction is needed when control elements are located
  around hubs ($q_1$) compared with the random distribution, whereas
  the required fraction to fight the mafia
  depends on the initial conditions for control elements attached to
  small nodes. Initial conditions are $m_0=c_0 =1/2$.}
\label{fig: ext police distributions}
\end{figure}

What is the system's behavior when control elements attach to specific targets as compared to the random case? The relation between the fraction of edges accommodating control elements and the fraction of nodes which are protected by those edges seems to be the key factor in the analysis of the problem. Figures \ref{fig: police distributions} and \ref{fig: ext police distributions} show the fraction of both populations and the extinction probability for the three distributions, $q_1$, $q_2$, and random, as a function of network homogeneity, $\gamma$.

The resulting fraction of controlled nodes, if control elements attach to large nodes, $q_1$, whose degree is close to the maximal value $k_M$, is smaller than in a random distribution,  
 $p/k_M < p/\left<k\right>$. Similarly, the effective fraction of protected nodes is larger than in the random case if control elements target sparsely connected nodes with the minimal degree $k_m$, $p/k_m > p/\left< k\right>$.  Thus, for a fixed police fraction, the population of mafiosi in equilibrium is larger for the $q_1$-distribution than for the random case, and this population is larger for the latter than for the $q_2$ distribution: $m(q_1) > m(\text{rand}) > m(q_2)$, as illustrated in Fig.~\ref{fig: police distributions}. The difference in the stationary population fractions induced by the specific distributions naturally increases with the heterogeneity of networks (small $\gamma$).

Fig.~\ref{fig: ext police distributions} illustrates that the fraction
of control elements at which extinction takes place increases if
the control elements concentrate around hubs ($q_1$). In this case, an
effective homogeneous structure is left unpoliced and, as we have
discussed before, even small fractions of mafiosi manage to invade a
citizens' population. In contrast, if the control elements preferentially surround
small nodes  ($q_2$), the unpoliced region consists of
highly connected nodes, whose dynamics is similar to
that  of the well-mixed population for the unpoliced asymmetric
model. In this case, the stationary state critically depends on the
initial conditions, so that we cannot draw any general conclusions
on the effect of the distribution on the system's behavior.
Whether a larger or smaller fraction of control elements than for the
random distribution is needed to drive the mafia to extinction depends
thus on the particular
initial conditions in the unpoliced areas.

\section{Conclusions and outlook}

We have proposed a novel way to model the influence of the neighborhood in social dynamics which accounts for the protection of alike peers proportional to their fraction. In this way, both coexistence of opinions and consensus on the neutral position (citizens, in the language used in the introduction) emerge as possible solutions in a natural way. We find that the dynamics of the model in structured societies drastically differs from that predicted for well-mixed populations on complete graphs. In the parameter window where the population dynamics exhibits bistability on complete graphs, we find coexistence between citizens and mafiosi on scale-free networks (and square lattices). This is traced back to the local character of the interaction which allows mafiosi to invade societies in structured societies with small neighborhood sizes. We have rationalized this effect employing both heuristic arguments on the population dynamics as well as an extended mean-field analysis, which accounts for the local network structure. Moreover, stochastic simulations for scale-free networks with increasing average degree $\langle k \rangle$ show that the coexistence region shrinks and the bistable regime is recovered in the limit of a complete graph~\cite{balbas2010diss}.

The presence of control elements (police) splits up the social structure into unconnected
subnetworks ruled by different dynamics.  Neutral individuals (citizens) populate regions influenced by the control elements (policed areas), while the remaining (unpoliced) areas develop the same dynamics as in the model without police, yielding a stationary state which depends on the network homogeneity. We have discussed the positive effect of this micro-phase separation on the population size of the minority group. In addition, we report a markedly sharp crossover from extinction of the minority to coexistence of species as a function of the heterogeneity of the underlying scale-free network. By preferential attachment
of the control elements to highly (sparsely) connected nodes, smaller (larger) effective fractions of the control elements emerge. Depending on the kind of nodes the control elements are preferentially attached to,  the areas free of external influence (no police) exhibit now a degree of heterogeneity different from that of the whole network. We have specifically investigated the role of two different distributions of the external elements on the given structure, showing that the attachment rules for the control elements determine the threshold fraction needed to bring the minority group (mafia) to extinction. 

We conclude with a comparison of the mafia model with other non-linear opinion model and generalizations thereof. The mafia model can be viewed as an extension of the Abrams-Strogatz (AS) model \cite{abrams2003} in the following sense. In the limit $d\to 0$ and $\phi\to 0$, where there are no more empty places in the mafia model, it reduces to a two-state reaction scheme $c\leftrightarrow m$ with transtion rates $\omega_{cm}= s_m m^2$ and $\omega_{mc}= s_c c^2$, i.e.\ an AS model with a low volatility of $a=2$. In the limit  of the asymmetric model, $s_c\to 0$, considered in this manuscript, the only stable fixed point is $(m,c)=(1,0)$, i.e. consensus on one of the two opinions as in the AS model. For general prestige values, $s_{m/c}\neq0$,  the coexistence fixed point is unstable and both consensus states become stable, i.e.\ the dynamics exhibits bistability.

These results highlight the importance of empty places (a third state) for the observed dynamics in the mafia model. Note, however, that the presence of empty places is only a necessary but not a sufficient condition for the emergence of bistability and coexistence. The functional form of the transition rates, i.e. the fact that citizens protect themselves, also matters in an important and interesting way. The transition rates of our model, e.g. $\omega_{cm} = s_m m (1-c)$, weaken the tendency to switch to the opposite state by a factor $(1-c)$ as compared to neutral voter-like models. Instead of these transition rates, we could have adopted an AS scheme to investigate the effect of peer support:  $\omega_{cm} = s_m m^a$ with $1<a<2$. However, these non-linear transition rates are qualitatively different from the mafia model: for a fixed fraction of empty sites, $\phi$, the rates in the mafia model are linear or non-linear depending on the relative fractions of species and empty sites, while they are always non-linear in the AS model.

Indeed, if one adopts transition rates like those of the AS model $\omega_{cm}=\sigma_m m^a$  and $\omega_{mc}=0$, i.e.\ generalizes the AS model to contain empty places, one does not find all the regimes of the mafia model. For $a=1$, there is a transition between two monostable regimes, extinction $m=0$ and coexistence at $\sigma = 1/(1+\beta)$. For $a=2$ there is a saddle-node bifurcation with a transition from a monostable regime with extinction of mafiosi to a bistable regime, where extinction and coexistence are stable fixed points. However, it does not show a monostable regime with coexistence of both species.  The mafia model exhibits richer behavior than its corresponding limits in terms of the generalized AS model, i.e.\ $a=1$ and $a=2$. In a sense, its dynamics combines all the regimes found for the two limiting cases of the AS model with empty sites: monostable regimes, with coexistence and extinction of mafiosi, and a bistable regime.

Our model still lacks the inclusion of the evolution of the network of contacts which takes place in actual societies.  Considering adaptive networks which also account for topological features is, in our view,  the most compelling extension of the model.

\section{Acknowledgements}
We  thank Heiko Hotz for the implementation of the code to generate scale-free networks. Financial support by the Deutsche Forschungsgemeinschaft through the SFB TR12 ``Symmetries and Universalities in Mesoscopic Systems" is gratefully acknowledged.

\appendix
\section{Local mean-field approximation}

A standard mean-field description as given by Eqs.~\eqref{eq: dotc}--\eqref{eq: dotphi} does not  properly capture the phase diagram of structured societes: the bistable regime of the mean-field theory largely becomes a coexistence regime in structured societies. For a well-mixed population, a standard mean-field theory which describes the
interaction of every agent with all others is expected to correctly predict
the stationary state of the system. However, we find from numerical simulations that in structured societies (networks and square lattices) with a one-to-all interaction scheme this standard mean-field behavior is not recovered even in the limit where individuals diffuse very strongly \cite{balbas2010diss}. The main reason is the one-to-all character of local interactions: while averaged frequencies properly reproduce the essence of pair interactions between two individuals (one-to-one), as in the voter model \cite{sood2008voter} or in the rock-paper-scissors game \cite{reichenbach2007mobility}, it fails to capture the dynamics of one-to-all interactions, due  to the discrete character of the neighborhood's composition. 

In the following we illustrate the relevance of the type of interaction for the phase behavior for square lattices without control elements. Here, the composition of the von Neumann
neighborhood cannot take arbitrary values, but instead the possible
fractions for both species and empty places are given by $0$, $1/4$,
$2/4$, $3/4$, $4/4$ with the constraint that $c+ m + \phi =1$. There
are thus fifteen possible neighborhoods $\Omega_i = (C_i, M_i,
\Phi_i)$ $=$ $(4,0,0)$, $(3,1,0)$, $\ldots$, $(0,0,4)$, with $C_i$, $M_i$,
and $\Phi_i$ denoting the total number of individuals of each species (in a given neighborhood) ranging from $0$ to $4$. Consequently, the transition rates
$\omega_{mc}$ and $\omega_{cm}$ are no longer continuous in structured societies.

\begin{figure}[h]
\centering
\resizebox{0.4\textwidth}{!}{%
\includegraphics{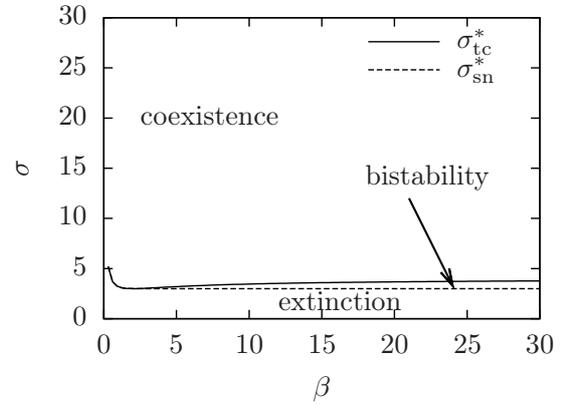}
}
\caption{Stability diagram for the asymmetric model in terms of the birth rate $\beta$ and the strength $\sigma$ within an extended mean-field approach. Note that the bistable region has largely shrunk as compared to the standard mean fiel shown in Fig.~\ref{fig: stability diagram}. The stability of the absorbing state $m^*=0$ is lost and, accordingly, the parameter space for coexistence of both species is strongly enlarged.}
\label{fig: local mf}
\end{figure}

The number of possible compositions for a neighborhood is finite in
networks and lattices. The smaller the neighborhood is, the smaller
the number of possible compositions and thus the further it is from
the asymptotically continuous case of the global population.  To
account for the finite number of available neighborhoods one has to
replace the averaged values of the species' frequencies entering the
transition rates, Eqs.~\eqref{eq: wcm} and \eqref{eq: wmc}, in a
mean-field description by a sum over all possible neighborhoods
$\Omega_i$ with their corresponding population fractions, weighted by
the probability $p_i$ for every neighborhood\footnote{To generalize
  the procedure for scale-free networks, one has to take the sum over
  the possible neighborhood's composition, after having considered a
  sum over the possible sizes of the vicinities, i.e.\ the degree $k$,
  weighted with the degree distribution.}:
\begin{eqnarray}
\omega_{cm} = \sigma m (1-c) &\longrightarrow &\sum_i
p_i\,\sigma\, \frac{ M_i }{4}\left(1-\frac{C_i}{4}\right),\label{eq:
  discretized omega cm}\\
\omega_{mc} = \sigma c (1-m) &\longrightarrow &\sum_i p_i \,
\sigma \,\frac{C_i}{4} \left(1-\frac{M_i}{4}\right).\label{eq:
  discretized omega mc}
\end{eqnarray}
In our extended mean-field approach we still assume that there  are no site correlations. Therefore, the probability to have $C_i$ citizens, $M_i$ mafiosi, and $\Phi_i$ empty places is that of drawing the specific combination out of a mixed sample with $Nc$ citizens, $N m$ mafiosi and $N\phi$ empty places. The corresponding probability for a given neighborhood  is thus given by the \emph{multinomial distribution}:
\begin{equation}
p_i = p(C_i,M_i,\Phi_i) =
\frac{(C_i+M_i+\Phi_i)!}{C_i!M_i!\Phi_i!}c^{C_i}m^{M_i}\phi^{\Phi_i}.\label{eq:
  multinomial probability}
\end{equation} 
Note that for a square lattice $C_i+M_i+\Phi_i=4$. Accounting for these probability distributions of possible neighborhoods in the generalized transition rates, Eqs.~\eqref{eq: discretized omega cm} and \eqref{eq: discretized omega mc}, we find that the fixed points of the ensuing extended mean-field theory are still an absorbing  and  two coexistence states as for the standard mean-field, but the functional form of the coexistence solutions now read:
\begin{eqnarray}
m^0 & = & 0,\\
m^{\pm} & = & \frac{(-2-\beta - \beta ^2 )\sigma \pm
  \sqrt{(1+\beta)^4\sigma (\sigma-3)}}{3\sigma (1+\beta)^2}.\label{eq: fixed point
  mixed lattice}
\end{eqnarray}
The lines at which the saddle-node, $\sigma^*_{\text{sn}} = 3$, and transcritical bifurcations, $\sigma_{\text{tc}}= 4(1+\beta)^2/(4\beta + \beta^2)$, occur are drastically shifted with respect to those for the standard mean-field theory. The parameter region of the bistable regime in largely reduced to a thin stripe, and the coexistence phase now covers a
much larger area in the parameter space, cf. Fig.~\ref{fig: local mf}.

The local mean-field analysis discussed here nicely captures the dynamics for structured populations, accounting for the loss of stability of the absorbing state for a large region of the parameter space in favour of coexistence in structures. The one-to-all interactions turns out to be a key feature of the mafia model to understand its reach dynamics, and in particular the hegemony of species coexistence in a variety of structures.

%\bibliographystyle{phjcp}
%\bibliography{ev-dyn}

\end{document}